\documentclass[prb,twocolumn,showpacs,preprintnumbers,amsmath,amssymb]{revtex4}

\usepackage{graphicx}% Include figure files
\usepackage{dcolumn}% Align table columns on decimal point
\usepackage{bm}% bold math
\usepackage{graphicx}
\usepackage{epsfig}
\include{graphic}

\begin{document}
\preprint{APS/123-QED}
\title{Far-infrared photo-conductivity of electrons in an array of nano-structured antidots}

\author{K. Bittkau, Ch. Menk, Ch. Heyn, D. Heitmann, and C. -M. Hu\footnote{Corresponding author. Email: hu@physnet.uni-hamburg.de}}
\affiliation{Institut f\"ur Angewandte Physik und Zentrum f\"ur
Mikrostrukturforschung, Universit\"at Hamburg, Jungiusstra\ss e
11, 20355 Hamburg, Germany}

\date{\today}

\begin{abstract}

We present far-infrared (FIR) photo-conductivity measurements for
a two-dimensional electron gas in an array of nano-structured
antidots. We detect, resistively and spectrally resolved, both the
magnetoplasmon and the edge-magnetoplasmon modes.
Temperature-dependent measurements demonstrates that both modes
contribute to the photo resistance by heating the electron gas
via resonant absorption of the FIR radiation. Influences of spin
effect and phonon bands on the collective excitations in the
antidot lattice are observed.

\end{abstract}


%72.25.Fe Optical creation of spin polarized carriers
%72.25.Rb Spin relaxation and scattering
%73.43.-f Quantum Hall effects
%73.43.Lp Collective excitations
%73.43.Qt Magnetoresistance
%73.50.Jt Galvanomagnetic and other magnetotransport effects (including thermomagnetic effects)
%73.50.Mx High-frequency effects; plasma effects
%73.50.Pz Photoconduction and photovoltaic effects
%73.63.Kv Quantum dots
%78.30.-j Infrared and Raman spectra
%78.67.Hc Quantum dots

\maketitle There has been recently growing interest to investigate
the photo resistance related with the elementary excitations of a
two-dimensional electron gas (2DEG) in semiconductor
heterostructures \cite{Zudov,Hu,Zehnder,Peralta,Jager,Ye}. For
example, under GHz radiation, the magneto-resistance of a 2DEG
shows an unexpected large oscillation whose period is determined
by the electron cyclotron resonance (CR) \cite{Zudov}. Using THz
radiation, plasmon of tunneling coupled bilayer 2DEGs is found to
contribute to the photo resistance in a unique way
\cite{Peralta}. Even the spin effects result in striking photo
resistance changes \cite{Hu,Zehnder}. Among others, the
spin-orbit interaction, which was about 80 years ago discovered by
the atomic spectroscopy and gave birth to the very concept of
\textit{spin}, has been found rather difficult to be spectrally
measured for the 2DEGs. The problem has been recently solved by
measuring the spin-flip excitation using the photo-conductivity
spectroscopy \cite{Hu}. Of particular interest is the high
sensitivity of photo-conductivity technique, which has the
potential to study unique elementary electronic excitations of
nano-structured semiconductors with only few units. As a first
step, very recently Jager \textit{et al.} \cite{Jager} and Ye
\textit{et al.} \cite{Ye} both studied the photo conductivity of
a 2DEG in an antidot array. While these nice experiments together
with the pioneer one of Vasiliadou \textit{et al.}
\cite{Vasiliadou} have shed light on the interesting photo
conductivity effect of a 2DEG in an antidot array, primary
questions like the role of the characteristic excitations of the
antidot array on its photo resistance are surprisingly left open,
which is the central subject of this work.

The elementary electronic excitations of a 2DEG in an antidot
array subjected to a perpendicular magnetic field $B$ are
dominated by a characterisic two-mode behavior with collective
excitations \cite{Mikhailov,Kern}. The upper mode $\omega^{+}$
approaches at large $B$ field the CR frequency $\omega_{c}
=eB/m^{\ast}$, which is determined by the electron effective mass
$m^{\ast}$. The lower one $\omega_{EMP}$, known as the
edge-magnetoplasmon (EMP) mode, is associated with the electrons
skipping around the depleted area with a radius $R$ formed by the
antidot potential. Using the modified-dipole and effective-medium
appoximations, the dispersion of these modes can be described by
\cite{Mikhailov}
\begin{equation}
1-\frac{(1-f)\omega_{0}^{2}}{\omega (\omega + \omega_{c})}-
\frac{f\omega_{0}^{2}}{\omega (\omega - \omega_{c})} = 0.
\end{equation}
%\begin{equation}
%1-\frac{1-f}{\frac{\omega}{\omega_{0}}(\frac{\omega}{\omega_{0}} +
%\frac{\omega_{c}}{\omega_{0}})} -
%\frac{f}{\frac{\omega}{\omega_{0}}(\frac{\omega}{\omega_{0}} -
%\frac{\omega_{c}}{\omega_{0}})} = 0.
%\end{equation}
Here $\omega_{0}$ is the frequency of the $\omega^{+}$ mode at $B$
= 0, depending on the antidot lattice period $a$, the 2DEG's
carrier density $N_{s}$, and its dielectric surrounding. The
geometrical filling factor $f = \pi R^{2}/a^{2}$ indicates the
portion of the 2D area where the electrons are depleted. These
modes have been well studied by far-infrared (FIR) transmission
spectroscopy \cite{Kern}, but have not been clearly identified in
photo-conductivity experiments \cite{Jager,Ye,Vasiliadou},
leaving the question open whether and how they might influence the
photo resistance of the 2DEG in an antidot array.

In this rapid communication, we report FIR photo-conductivity
experimental results obtained for a 2DEG in an antidot array. We
find clearly that both the $\omega^{+}$ and $\omega_{EMP}$ modes
contribute to the photo resistance by heating the electron gas via
resonant absorption of the FIR radiation. In addition, we present
interesting results indicating the influences of spin and
electron-phonon interaction in the antidot lattice.

Our sample is an inverted-doped InAs step quantum well with 40 nm
In$_{0.75}$Al$_{0.25}$As cap layer. The step quantum well is
composed of 13.5 nm In$_{0.75}$Ga$_{0.25}$As, an inserted 4 nm
InAs channel, and a 2.5-nm-thick In$_{0.75}$Ga$_{0.25}$As layer.
Underneath the quantum well is a 5 nm spacer layer of
In$_{0.75}$Al$_{0.25}$As on top of a 7-nm-wide Si-doped
In$_{0.75}$Al$_{0.25}$As layer. The sample is grown by molecular
beam epitaxy on a buffering multilayer accommodating the lattice
mismatch to the semi-insulating GaAs substrate. A self-consistant
Schr\"{o}dinger-Poisson calculation shows that the 2DEG is about
55 nm below the surface, mainly confined in the narrow InAs
channel \cite{Richter}. Figure 1 shows a sketch of our sample with
antidots. An extreme long 2DEG Hall bar with a channel width of
$W$ = 40 $\mu$m and a total length $L$ of about 10 cm was defined
by chemical wet etching, which contains the antidot array with a
period of $a$ = 800 nm. The holes with a geometric diameter of
about 200 nm were defined by holography and chemical wet etching.
The 2DEG channel runs meandering in a square of 4 $\times$ 4
mm$^{2}$. The extremely large $L/W$ ratio enhances the
sensitivity of our measurement. With the antidots, the carrier
density $N_{s}$ and mobility $\mu$ at 1.5 K were determined by
Shubnikov-de Haas measurement to be 6.01 $\times$ 10$^{11}$
cm$^{-2}$ and 62 000 cm$^{2}$/Vs, respectively, reduced compared
with those of the corresponding unpatterned sample \cite{Hu} of
6.66 $\times$ 10$^{11}$ cm$^{-2}$ and 150 000 cm$^{2}$/Vs. Ohmic
contacts were made by depositing AuGe alloy followed by annealing.

Our experiment was performed by applying a DC current of 9 $\mu$A
to the Hall bar and measuring the changes of the voltage drop
caused by FIR radiation. At fixed magnetic fields, the broadband
FIR radiation was modulated by the Michelson interferometer of a
Fourier transform spectrometer. Using the sample itself as the
detector, the corresponding change in the voltage drop of the
sample was AC coupled to a broadband preamplifier and recorded as
an interferogram, which was Fourier transformed to get the
photo-conductivity spectrum. The sample was mounted in a He
cryostat with a superconducting solenoid. All data reported here
were obtained in Faraday geometry.

\begin{figure}
\begin{center}
%\hskip -0.7cm
\epsfig{file=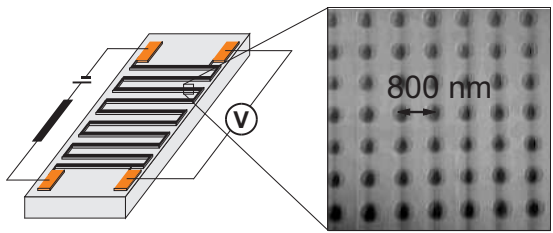,height=3cm} \caption{Schematic bias circuit
and sample structure showing the long Hall bar with ohmic
contacts. The zoom-in part is an atomic force micrograph of the
antidot array.} \label{fig_Fig1}
\end{center}
\end{figure}

Figure 2 shows typical FIR photo-conductivity spectra measured at
different $B$ fields and temperatures. Different beam splitters of
the spectrometer are used to optimize the measurement for
resonances lying in different frequency regimes. As shown in
Fig.~2(a) at the low magnetic field of $B$ = 3.2 T, two
resonances are clearly observed. By increasing the $B$ field to
6.4 T, the resonance at the lower energy has a slight red shift
and gets weaker, while the higher-energy resonance shows a
significant blue shift and dominates the spectrum. Two additional
weak resonances are observed, one appears as a shoulder of the
dominant resonance and the other lies at about 285 cm$^{-1}$,
which are indicated by a thick and thin arrow, respectively. By
further increasing the $B$ field to 9.6 T, the dominant resonance
splits into multi-peaks, while the weak one at 285 cm$^{-1}$
gains resonance strength. The resonance at the low energy
disappears at large $B$ fields. In Fig.~2(b) we plot the spectra
measured at $B$ = 3.2 T and at different temperatures. In this
case the DC current was reduced to 180 nA to avoid heating the
2DEG by the current. The observed resonances are found extremely
temperature sensitive, with their amplitudes decreasing quickly
by only slightly increasing the temperature.

\begin{figure}
\begin{center}
\epsfig{file=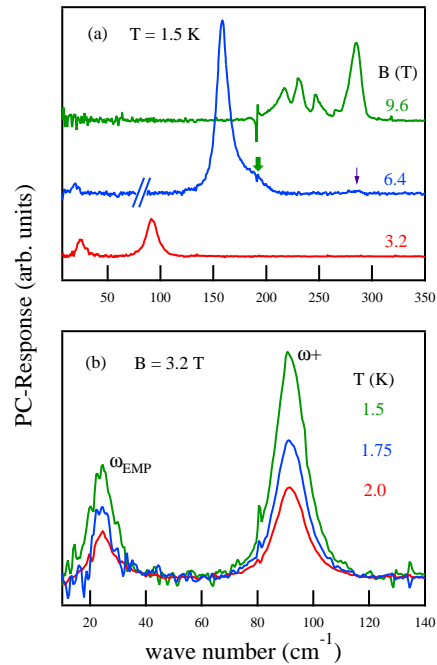,height=9cm} \caption{(color online). FIR
photo-conductivity spectra measured at (a) $T$ = 1.5 K for three
different magnetic fields and (b) $B$ = 3.2 T for three different
temperatures. Arrows indicate weak resonances described in the
text. Spectra in (a) are vertically offset for clarity.}
\label{fig_Fig2}
\end{center}
\end{figure}

In Fig.~3(a) we plot the $B$-field dispersion of these resonances.
Also shown is the magnetoresistance $R_{xx}$ measured without FIR
radiation using the standard lock-in technique, which allows us to
determine the 2DEG filling factors $\nu =N_{s}h/eB$. The major
resonances can be nicely fit (solid curves) using eq.~(1) with
three fitting parameters $\omega_{0}$ = 70.2 $cm^{-1}$, $f$ =
0.17, and $m^{\ast}$ = 0.039 $m_{e}$. Within the fitting
accuracy, the obtained effective mass value is equal to that
directly measured from CR on the unpatterned sample from the same
wafer \cite{Hu}. We therefore identify them as the two
characteristic antidot collective modes $\omega^{+}$ and
$\omega_{EMP}$. The relative strength of $\omega^{+}$ over
$\omega_{EMP}$ mode increases with increasing $B$ field, in
accordance with the theory \cite{Mikhailov}. By comparing the
spectra with that obtained on the unpatterned sample, we further
identify the weak resonance marked by the thick arrow in Fig.~2(a)
as the collective spin-flip excitation \cite{Hu}. With the
antidot lattice, the collective spin-flip excitation gets broader
and appears as a shoulder of the $\omega^{+}$ resonance. Spin-flip
excitation in an antidot array has neither been studied
experimentally nor been theoretically investigated. Here we have
assumed that at large $B$ fields, the spin-flip excitation in the
antidot array approaches that in an unpatterned 2DEG, just like
the $\omega^{+}$ mode approaches the CR \cite{Mikhailov}. For
comparison, we plot in Fig.~3(a) the calculated dispersion for
the 2DEG spin-flip excitation neglecting both the many-body
correction and the antidot potential, using the spin-orbit
coupling parameter of $\alpha = 2.38\times 10^{-11}$ eVm and the
Land\'{e} $g$-factor of $g$ = -8.7 determined for the
unpatterened sample \cite{Hu}. Within the $B$-field range of 6 to
7 T where we can observe the spin-flip excitation, influence of
the antidot potential on its resonance frequency is found small.

\begin{figure}
%\begin{center}
\hskip -0.7cm\epsfig{file=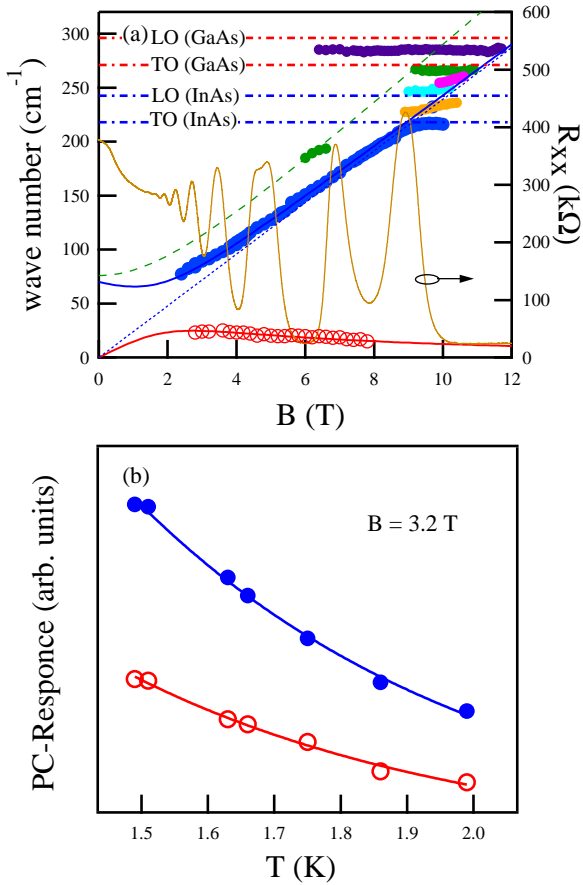,height=12cm} \caption{(color
online). (a) Magnetic-field dispersions for resonances measured
at $T$ = 1.5 K and magnetoresistance $R_{xx}$ measured without
FIR radiation. The solid curves are fits for $\omega^{+}$ (solid
circles) and $\omega_{EMP}$ (open circles) antidot modes using
Eq.~(1). The dotted line and dashed curve are calculated for CR
and spin-flip excitation, respectively, using the effective mass
of 0.039 m$_{e}$, the spin-orbit coupling parameter of $\alpha =
2.38\times 10^{-11}$ eVm and a Land\'{e} $g$-factor of $g$ =
-8.7. Dash-dotted lines indicate the optical phonon energies of
InAs and GaAs. (b) Temperature dependence of the resonance
strength for the $\omega^{+}$ (solid circles) and $\omega_{EMP}$
(open circles) antidot modes measured at $B$ = 3.2 T. The curves
in (b) are guide to eyes.} \label{fig_Fig3}
%\end{center}
\end{figure}

Observing the spin-flip excitation in the antidot lattice
demonstrates the high sensitivity advantage of the
photo-conductivity spectroscopy. Another advantage of the
technique is that we can measure resonances within the
reststrahlen bands which are prohibited for transmission
spectroscopy. In our sample, both the InAs quantum well and the
GaAs substrate are polar semiconductors with reststrahlen bands
between their TO and LO phonon frequencies. Besides, there are
two phonon bands for both In$_{0.75}$Ga$_{0.25}$As and
In$_{0.75}$Al$_{0.25}$As layers, as well as the interface phonons
near each interfaces. For brevity, we plot in Fig.~3(a) the bulk
phonon frequency of GaAs and InAs. Apparently, the splitting of
the $\omega^{+}$ mode at large $B$ fields is caused by the
influence of phonons of our sample. Experimentally, by increasing
the $B$ field, we find that the resonance at about 285 cm$^{-1}$
indicated by the thin arrow in Fig.~2(a) stays within the
reststrahlen band of GaAs substrate but gains resonance strength.
While the splitting of the $\omega^{+}$ mode shows anti-crossing
behavior centered at about 220 and 240 cm$^{-1}$, near the TO
(218 cm$^{-1}$) and LO (242.5 cm$^{-1}$) phonon frequency of InAs,
respectively. At the moment, we cannot explain these resonances
in the reststrahlen bands regime. We note that the influence of
phonons on electronic excitations of a nano-structured 2DEG in the
reststrahlen band regime is a rather sophisticated problem, with
the combined nature of the optical effect \cite{Ziesmann}, band
nonparabolicity, electron-electron and electron-phonon
interaction \cite{Hu95,Wu}. It remains a controversial subject
\cite{Poulter}. However, the rich spectral features observed in
our experiment provide systematic data that is essential to
establish a clear theoretical picture.

Our data demonstrates clearly that both $\omega^{+}$ and
$\omega_{EMP}$ antidot modes contribute to the photo resistance.
In contrast to that studied by the transmission spectroscopy, the
resonance strengths of both modes measured by photo-conductivity
spectroscopy are extremely temperature sensitive. As shown in
Fig.~2(b) the resonance strengths of both modes decrease quickly
by increasing the temperature from 1.5 to 2 K, during which the
half width of the resonances does not change much. In Fig.~3(b),
we plot the temperature dependence of the resonance strength for
both modes measured at $B$ = 3.2 T. Such an extremely sensitive
temperature dependence can be qualitatively explained by the
bolometric effect \cite{Zehnder,Neppl}, where the resonant
absorption of the FIR radiation effectively heats the electron
gas and hence changes its resistance. In the low temperature
limit $k_{B}T << E_{F}$, the heat capacity of the 2DEG is given
by \cite{Wang}
\begin{equation}
C_{e}=\pi^{2}k_{B}^{2}T\mathcal{D}\mathrm{(E_{F},B)/3},
\end{equation}
proportional to the temperature $T$ and the density of states
$\mathcal{D}\mathrm{(E_{F},B)}$ of the 2DEG at the Fermi energy
$E_{F}$. Bolometric effect caused by electron heating is therefore
more pronounced at lower temperature where $C_{e}$ is smaller.

It is intriguing to compare our results with other
photo-conductivity experiments on the 2DEG in an antidot array. In
the early work of Vasiliadou \textit{et al.} \cite{Vasiliadou},
instead of a broad band FIR source, microwave generator was used
to investigate the commensurability effects. Photo-conductivity
experiment was performed by fixing the microwave frequency while
sweeping the magnetic field. Characteristic dispersions for the
antidot collective modes were not investigated due to the limited
available microwave frequencies. And temperature dependence was
not easy to be studied because of the superposed nonresonant
background which originates from heating of the whole sample. In
an improved photo-conductivity spectroscopy experiment recently
performed by Jager \textit{et al.} \cite{Jager}, the 2DEG is at a
distance of 37 nm below the sample surface and the antidots are
written by $e$-beam lithography and transferred into the 2DEG by
shallow (with only about 6 nm) wet etching. Therefore, instead of
$\omega^{+}$ and $\omega_{EMP}$ modes, CR and a magnetoplasmon
mode were simultaneously observed, which is typical for a weak
modulated 2DEG instead of the 2DEG with antidot confinement. The
most recent experiment was performed by Ye \textit{et al.}
\cite{Ye} using a significantly improved microwave technique with
transmission lines to study both the photo-conductivity and
microwave transmission of a 2DEG in an antidot array. Among other
interesting results, they observed a broad peak below Landau
filling one in the microwave conductivity measured by sweeping
the $B$ field. On the peak, unlike the DC-limit conductivity, the
microwave conductivity was found extremely temperature sensitive,
decreasing with increasing temperature. As the possible origins,
antidot edge excitations of fractional quantum Hall effect states
associated with either the chiral Luttinger liquids \cite{Ye} or
edge reconstruction \cite{Wan} have been discussed. Our data
provides additional insight. As shown in Fig.~3(b), we have
demonstrated that the contribution of the antidot
edge-magnetoplasmon to the photo resistance has similar sensitive
temperature dependence due to its bolometric nature. Besides,
based on the flat $B$-field dispersion of the edge-magnetoplasmon
shown in Fig.~3(a), photo resistance caused by exciting the
edge-magnetoplasmon at a fixed frequency would show broad
structures, superposed by nonresonant background due to the
change of the photo-conductivity sensitivity by sweeping the $B$
field. However, we would like to emphasize that whether the
edge-magnetoplasmon indeed plays a role in the nice experiment of
Ye \textit{et al.} \cite{Ye} depends on two major questions: the
first one is how does the photo resistance influence the
microwave transmission in their experiment; and the second one is
how large is the edge-magnetoplasmon frequency in their sample.
We note that while the first question is not easy to answer due
to the complicated microwave technique. The second question can
be clarified by performing photo-conductivity spectroscopy
experiment as we describe in this paper. By measuring the
dispersion of the $\omega^{+}$ mode in the FIR regime, the
geometric filling factor $f$ that depends on the depletion area
of the antidot can be determined, which can be used to estimate
the $\omega_{EMP}$ mode frequency \cite{EMP}. In the sample of Ye
\textit{et al.} \cite{Ye} the later lies in the microwave regime
and is rather difficult to be directly measured.

In summary, we have performed FIR photo-conductivity spectroscopy
experiment on a 2DEG in an array of antidot. We find that both the
magnetoplasmon ($\omega^{+}$) and the edge-magnetoplasmon
($\omega_{EMP}$) modes of the antidot contribute to the photo
resistance, which is extremely temperature sensitive due to the
bolometric nature. We observe the influence of phonon bands on the
$\omega^{+}$ mode and a spin-flip excitation mode in the antidots.

This work is supported by the BMBF through project 01BM905 and
the DFG through SFB 508. We thank Axel Lorke and Peide Ye for
helpful discussions about the experiments of ref. 5 and 6,
respectively.


\begin{thebibliography}{20}
\bibitem{EMP}
One might need also estimating the half width of the
$\omega_{EMP}$ mode (of the order of 100 GHz in our sample), since
exciting the tail of the $\omega_{EMP}$ mode could also cause a
$B$-field dependent photo resistance.
\end{thebibliography}
\end{document}